\documentclass[apj]{emulateapj}
\usepackage{natbib}
\usepackage{graphicx}
\bibpunct{(}{)}{;}{a}{}{,}

\newcommand\los{LOS}

\begin{document}

   \title{SUNRISE/IMaX observations of convectively driven vortex flows in the Sun}
   \author{
 	J.~A.~Bonet\altaffilmark{1,2},
 	I.~M\'arquez\altaffilmark{1,3},
 	J.~S\'anchez Almeida\altaffilmark{1,2},
 	J.~Palacios\altaffilmark{4},
 	V.~Mart\'\i nez Pillet\altaffilmark{1},
 	S.~K.~Solanki\altaffilmark{5,6},
 	J.~C. del~Toro~Iniesta\altaffilmark{7},
 	V.~Domingo\altaffilmark{4},
  	T.~Berkefeld\altaffilmark{8},
 	W.~Schmidt\altaffilmark{8},
 	A.~Gandorfer\altaffilmark{5},
 	P.~Barthol\altaffilmark{5},
	and
 	M. Kn\"olker\altaffilmark{9}
}
    \altaffiltext{1}{Instituto de Astrof\'\i sica de Canarias, 
              E-38205 La Laguna, Tenerife, Spain}
    \altaffiltext{2}{Departamento de Astrof\'\i sica, Universidad de La Laguna,
        E- 38071 La Laguna, Tenerife, Spain}
    \altaffiltext{3}{Departamento de An\'alisis Matem\'atico, 
        Universidad de La Laguna, E-38271  La Laguna, Tenerife, Spain}
    \altaffiltext{4}{Laboratorio de Procesado de Im\'agenes, 
        Universidad de Valencia, E-46980
        Paterna, Valencia, Spain}
     \altaffiltext{5}{Max Planck Institut f\"ur Sonnensystemforschung, 
        Max Planck Strasse 2, Katlenburg-Lindau, 37191, Germany}
     \altaffiltext{6}{School of Space Research, Kyung Hee University, 
        Yongin, Gyeongg 446-701, Republic of Korea}
     \altaffiltext{7}{Instituto de Astrof\'\i sica de Andaluc\'\i a, 
        Camino Bajo Huetor 50, 18008, Granada, Spain}
     \altaffiltext{8}{Kiepenheuer-Institut f\"ur Sonnenphysik, 
        Sch\"oneckstr. 6, D-79110, Freiburg, Germany}
     \altaffiltext{9}{High Altitude Observatory, National Center for 
        Atmospheric Research, P.O. Box 3000, Boulder CO 80307-3000, USA. 
        (NCAR is sponsored by the National Science Foundation)}
\begin{abstract}
We characterize the observational properties of the 
convectively driven vortex flows recently discovered
on the quiet Sun, using magnetograms, Dopplergrams and images
obtained with the 1-m balloon-borne {\sc Sunrise} telescope
. By visual inspection of time series, 
we find some $3.1\times 10^{-3}$~vortices~Mm$^{-2}$\,min$^{-1}$
, which is a factor of $\sim$1.7 larger than previous estimates. The mean duration of the individual events turns out to
be 7.9\,min, with a standard deviation of 3.2\,min. 
In addition, we find several 
events appearing at the same locations along the 
duration of the time series (31.6\,min). Such 
recurrent vortices show up in the proper motion flow field map 
averaged over the time series.
The typical vertical vorticities are 
$\lesssim 6\times$10$^{-3}\,$sec$^{-1}$, 
which corresponds to a period of rotation of 
some 35\,min.
The vortices show a preferred counterclockwise
sense of rotation, which we conjecture may have to do with the
preferred vorticity impinged by the solar differential rotation.
\end{abstract}


\slugcomment{ver0}

\shorttitle{SUNRISE/IMaX vortices}
\shortauthors{Bonet et al.}

%
\section{Introduction}\label{intro}

Solar surface convection is driven by localized downdrafts
that collect the cold plasma returning to the solar interior 
after releasing internal energy \citep[e.g.][]{spr90,stei98,ras98}.
Angular momentum conservation forces the plasma to
spin up as it approaches the sinkhole, and
vortices are formed at the downdrafts.
Such convectively driven vortices were theoretically
predicted and sought for long \citep[e.g.,][]{spr90,bal98}, 
but their observational discovery is fairly recent 
\citep[][with the well-known earlier detection of a single
large whirlpool by \citeauthor{bra88}~\citeyear{bra88}]
{bon08,wed09,bal10,goo10}.
In addition to supporting 
numerical models of solar surface convection, 
the photospheric vortices may be of importance as
heating sources for the outer solar atmosphere. 
The downdrafts not only advect vorticity but also 
magnetic fields, which are intensified to kG field 
strengths in and around them. Buoyancy and the 
vertical geometry of the downdraft tend to align the 
magnetic field lines with the vertical, so that the 
spinning motions at photospheric levels can be 
propagated upward using the field lines as guides 
\citep[e.g.,][]{cho93,zir93,bal98}.  Waves thus 
excited transport photospheric energy that can be 
deposited in higher layers of the atmosphere. 
Moreover, downdrafts often 
trap structures of mixed polarity, so that the swirling 
motions wind up opposite polarity field lines, 
facilitating magnetic reconnection and the ensuing 
energy release.

These convectively driven vortex flows are a recently discovered 
phenomenon poorly characterized
from an observational point of view.
So far, we only know that the vortices are quite 
common, have no preferred sense of rotation 
at the solar equator, and 
last (at least) minutes \citep[][]{bon08}. They are 
also visible in the chromosphere, where they seem
to be associated with significant blueshifts \citep[][]{wed09}.
Most of them are small-scale \citep[$\lesssim 0.5$\,Mm][]{bon08},
but some have a much larger radius of influence
\citep[up to 20\,Mm, ][]{att09,bra88}. 
Lifetimes can be longer than 20\,min, and 
several observables (such as circular polarization and 
G-band intensity) simultaneously indicate the 
presence of vortical motions \citep[][]{bal10}. 
In terms of global properties rather than individual eddies, 
the vertical vorticity inferred from proper motions 
seems to be higher in downflow regions, suggesting excess 
vorticity in intergranular lanes \citep[][]{wan95b,pot05}.
\citet{nise03} searched for evidence of vorticity 
in the motions of isolated G-band bright points.

{\sc Sunrise} is a 1-m balloon-borne solar telescope 
\citep[][]{bar10} which, together with its 
Imaging Magnetograph eXperiment
\citep[IMaX, ][]{mar10}, provides
time series with state-of-the-art high spatial 
resolution images and magnetograms. 
The dataset is ideal for a systematic characterization
of the poorly known physical properties of the vortices. 
Thus this 
Letter presents a comprehensive observational 
characterization  of the vortices in the photosphere. 
The actual data set and the procedure to detect vortices 
are described in \S~\ref{description}.
The main results are summarized in \S~\ref{results}. 
Based on such results, we compare the observed properties 
with the predictions of the numerical simulations of 
solar surface convection (\S~\ref{discussion}).

%
\section{Observation and analysis procedure}\label{description}

The data were gathered with IMaX near the solar disk center on June 9, 2009 (UT
01:31-02:02; although the exact location is not known) 
during the first science flight of {\sc Sunrise} \citep{bar10,sol1}.  The
 IMaX magnetograph uses a LiNbO$_3$ etalon operating in double
pass, liquid crystal variable retarders as the polarization modulator, and a
beam splitter as the polarization analyzer.  
We use here data recorded in the so-called V5-6
observing mode \citep[see][]{mar10} where images of the four Stokes
parameters were taken at five wavelengths along the profile of the
magnetic-sensitive line Fe~{\sc i\,}$\lambda$5250.2 \AA\ ($\pm 80, \pm
40~$m\AA\ from line center, plus continuum at +227\,m\AA). 
After the science observing run a calibration set consisting of 30 in-focus and 
out-of-focus image-pairs was recorded for post-facto retrieval
of the point spread function (PSF) using phase diversity
\citep[][]{gon82,pax96}. 
The science images were reconstructed by deconvolution using a modified Wiener 
filter and the calibrated PSF of the optical system.
 IMaX provides 85\,m\AA\ spectral resolution
and between 0\farcs15 and 0\farcs18 angular resolution in the reconstructed images. 
Dopplergrams and
magnetograms are derived from the Stokes parameters by using the approach
described in \citep[][]{mar10}.
 All in all, the reduction procedure renders time series of
images, magnetograms, and Dopplergrams with a  cadence of 33\,sec, a
spatial sampling of 0\farcs055, and an effective field-of-view (FOV) of
45\arcsec$\times$45\arcsec . As inferred from the standard deviation of the
polarization signals at the continuum wavelength, the circular polarization
noise is $5\times 10^{-4}$ in units of the continuum  intensity.  The
observing material analyzed here consists of a time series lasting 31.6 min.
Movies were generated after rigid alignment and $p$-mode subsonic filtering
\citep{tit89}.

\citet{bon08} found the small-scale vortex flows
by visual feature tracking of magnetic G-band bright 
points (BPs). The same downdrafts producing vortices 
also advect and concentrate magnetic flux (see \S~\ref{intro}), 
which often appears as BPs when the field strength is in the 
kG regime \citep[see, e.g.,][and references therein]{san04a}.
As Bonet and colleagues acknowledge, the technique is
rather limited since whirlpools without BPs are expected,
and they escape from detection. Taking advantage of the 
combined high spatial resolution and high polarimetric sensitivity 
 of IMaX/{\sc Sunrise} data, we tried to detect and study vortices in 
longitudinal magnetograms, which are sensitive not only 
to kGs fields, but to plasmas with the full range of field strengths. In addition, 
we broaden the study using continuum intensity, line minimum intensity, 
line-of-sight (LOS) velocity, and line width  (the last three parameters obtained 
from a Gaussian fit to the five sampled wavelengths). 
Individual vortices are identified and characterized through the
following steps:

(1) The detection is based on a visual inspection of
longitudinal-magnetogram movies. Playing back and forth these movies,  
we identify those locations and time intervals where structures 
seem to rotate. 

\begin{figure*}
 \includegraphics[width=0.9\textwidth,angle=0]{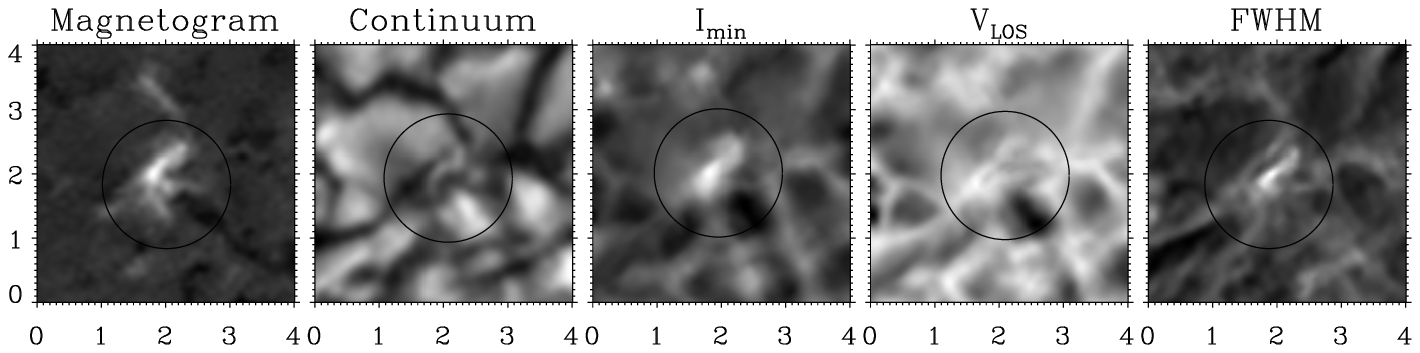}
\caption{
A particular vortex as reflected in different
observational parameters -- from left to right: magnetogram, 
continuum intensity, line core intensity,
LOS velocity, and line width. The panels show the average 
along the life-time of the event ($\sim 6.7$ \,min), for every parameter.
Horizontal scales are given in Mm. The 1 -Mm radius circles 
centered in the sinkhole are included for reference.
}
\label{fig1}
\end{figure*}

(2) Once a vortex candidate is thus located, 
it is isolated in a  
5\farcs5\,$\times$\,5\farcs5 sub-field, where the corresponding sub-fields 
in the other four physical parameters are visually inspected for 
swirling motions (see Fig.~\ref{fig1}).

\begin{figure*}
 \includegraphics[width=0.9\textwidth,angle=0]{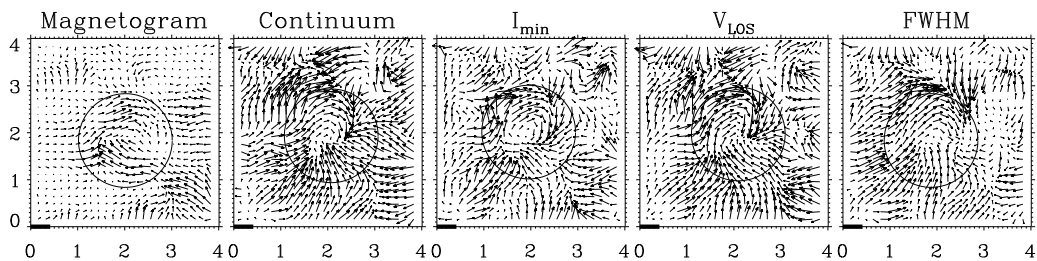}
\caption{
Horizontal velocity maps derived by LCT, 
from the proper motions 
of the parameters shown in Fig.1. The velocities are averaged 
over the life-time of the event ($\sim 6.7$ \,min). Horizontal 
scales are given in Mm. 
The length of the black bar at coordinates (0,0) corresponds 
to 1.8 km\,s$^{-1}$.} The 1 -Mm radius circles 
centered in the sinkhole are included for reference.
\label{fig2}
\end{figure*}

(3) Horizontal velocity maps of the event in all the five parameters
are created from proper motions. The horizontal motions are measured in these reduced 
sub-fields  employing the local correlation tracking (LCT) algorithm 
of \citet[][]{nov88}, as implemented by \citet[][]{mol94},  and
with a Gaussian tracking window of about 0\farcs4 FWHM. In order to help the algorithm, 
the original images are interpolated in time and space so as to have a pixel of
0\farcs028 and a cadence of 11\,sec.  The horizontal velocities obtained by comparing successive images  are time averaged the duration of the event. 
Examples of such velocity maps are shown in Fig.~\ref{fig2}. 
The size of the tracking window was chosen as a trade off
to be large enough for the LCT algorithm to have structures to 
track, yet small enough to minimize the presence of several structures 
with different velocities. 
If the velocity maps do not show a regular closed shape in at
least two physical parameters,
then we discard the vortex candidate and start from 
step 1.

\begin{figure*}
 \includegraphics[width=0.9\textwidth,angle=0]{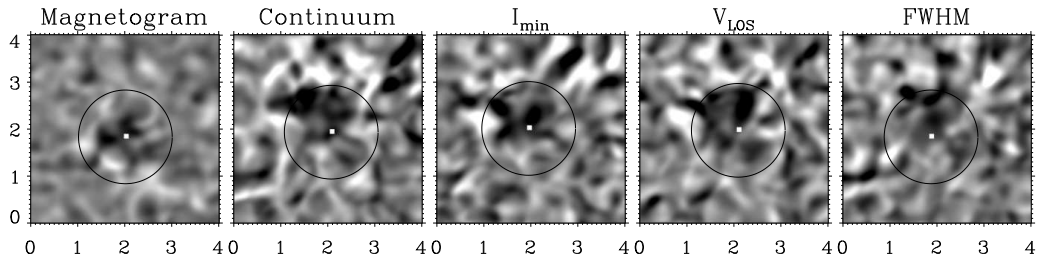}
\caption{
Maps of vertical vorticity corresponding to the LCT horizontal 
velocities for the parameters shown in Fig.1. The white points 
indicate the final position of the corks in the 
cork movies (see text). Horizontal 
scales are given in Mm. The 1 -Mm radius circles 
centered in the sinkhole are included for reference. Vorticities 
are represented in the range $\pm 6\times 10^{-3}$ s$^{-1}$, using a common 
 grayscale in all the panels.
}
\label{fig3}
\end{figure*}

\begin{figure*}
 \includegraphics[width=0.9\textwidth,angle=0]{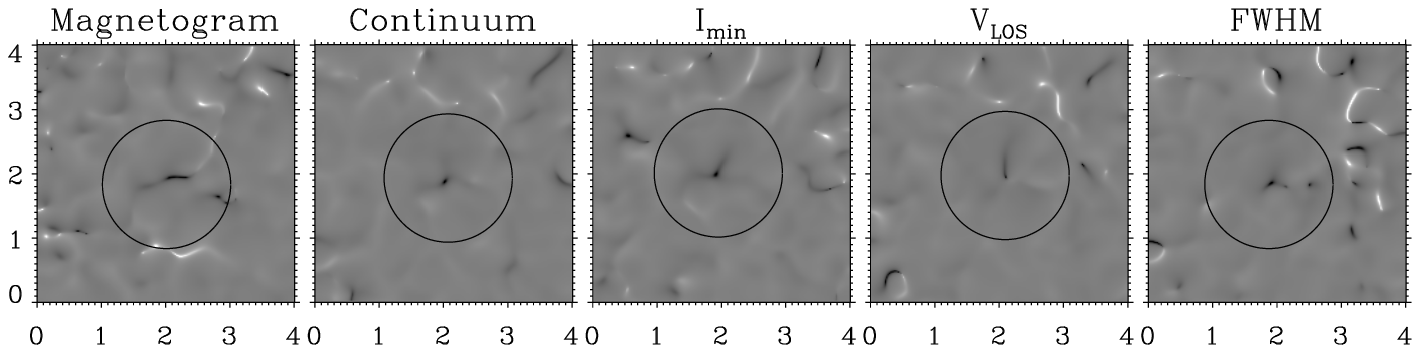}
\caption{
Curvature maps derived from the LCT horizontal 
velocities for the parameters shown in Fig.1.
Horizontal scales are given in Mm.
The 1 -Mm radius circles centered in the sinkhole
are included for reference. Note how the curvatures 
clearly show the position of the sinkhole as an intense point-like
feature. Curvatures are represented in the range 
$\pm 5\times 10^{-2}$ km$^{-1}$, using a common 
 grayscale in all the panels.
}
\label{fig4}
\end{figure*}

(4) We compute the vertical vorticity, the divergence, and 
the curvature corresponding to the LCT horizontal velocities. Given
the velocity {\bf U}, the vertical vorticity
$(\nabla\times {\bf U})_z$ can be interpreted in terms of 
the local angular velocity since a plasma in pure rotational 
motion has $(\nabla\times {\bf U})_z= 2 w$, with $w$ the angular 
velocity. Similarly, the curvature of such motion is  
$\kappa = {{1}\over{2}}\,(\nabla\times {\bf U})_z\,|{\bf U}|^{-1}$, 
with $\kappa^{-1}$ the radius of curvature.
Examples of vorticity and curvature maps
are shown in Figs.~\ref{fig3},~\ref{fig4}.
%

%
 
(5) Using the LCT horizontal velocities,
we track the evolution of passively advected
tracers ({\em corks}) spread out all over 
the sub-field \citep[][]{yi92}. 
As time goes by, the corks end up in the sinkhole, 
revealing the position of the sink according 
to different physical parameters 
(see the white points in Fig.~\ref{fig3}).
If the sinkhole positions inferred from the different
physical parameters do not agree within 0\farcs 4,
then the vortex is discarded and we return to step 1.

We find the curvature maps to be an efficient complementary tool 
for vortex center detection. In most cases these maps show up the 
sinkholes as conspicuous point-like features 
(see Fig.~\ref{fig4}) with positions that agree well 
with the centers determined as the final destination of the 
corks in the corresponding movie 
(c.f. Figs.~\ref{fig3},~\ref{fig4}).

The above list outlines the general procedure,
but we do not disregard casual detections, e.g., 
when a second vortex was observed in any
of the subfields corresponding to another
vortex. Thus we find by chance some vortices 
which do not show up in the magnetogram signals.
In addition to the LCT velocity maps of the 
individual vortices, we also computed the flow field 
for the full FOV during the full duration of the 
sequence. Isolated point-like features in the 
corresponding curvature map suggest
the presence of vortices, and their existence
motivates further inspection of 
magnetograms for swirling motions.
As one can image from this cumbersome procedure,
the FOV has been unequally searched. We focused on those 
regions were the magnetograph signals were largest, 
so that the {\em effective} FOV of our research is 
only 28\farcs 5 $\times$ 28\farcs 5. This area is used 
in the estimates below unless otherwise stated.

Two animations, one showing the event in Figs.~\ref{fig1}-\ref{fig4} 
and another with a different one, are given in the on-line material 
in the electronic edition of the Journal.

%

%
\section{Results}\label{results}

Following the procedure outlined above, we detected 
42 vortices with proper velocity maps.
They imply a space-time-density of 
$d\simeq 3.1\times 10^{-3}$~vortices~Mm$^{-2}$\,min$^{-1}$.
In addition, 31 structures showing vortical 
motion were discarded because 
they did not fullfill our strict selection criteria. 
 The selected events are used here to characterize the 
observational properties of vortices.
  
%
We have assigned a duration to each vortex, computed as the time
interval in which the vortex motions are clearest. 
 These durations span from 5\,min to 20\,min
with a mean of $\tau$ $\simeq$ 7.9\,min and a standard deviation of 3.2\,min.
The interpretation of these intervals as life-times is not 
devoid of uncertainty, though. Often we shorten the interval to 
assure a most pure swirling motion. In addition,
some vortices appear in the mean flow field corresponding 
to the full time series, indicating that they probably last 
longer than the time span of the series itself. 
These long lasting vortices often involve a complex 
 behavior: several short-lived vortices appear and disappear in the 
same location, giving rise to recurrent vortices. 
The position of their vortex centers may be static 
or drift with time. The recurrent vortices
 may or may not keep 
the same sense of rotation. In the latter case, however, 
we cannot consider the vortices to be strictly recurrent. 
We even find cases where the 
presence of a vortex is hinted as a clear pointlike feature in 
the curvature image, but we failed to identify any  
vortex at that position during the sequence. 





The largest surprise of our analysis is the 
finding of a preferred sense of rotation: 
27 counterclockwise vs. 15 clockwise.
This is a big difference with respect to \citet{bon08},
where the two senses of rotation were 
observed equally. We analyze this issue in \S~\ref{discussion}.

\begin{figure}
 \includegraphics[width=0.5\textwidth]{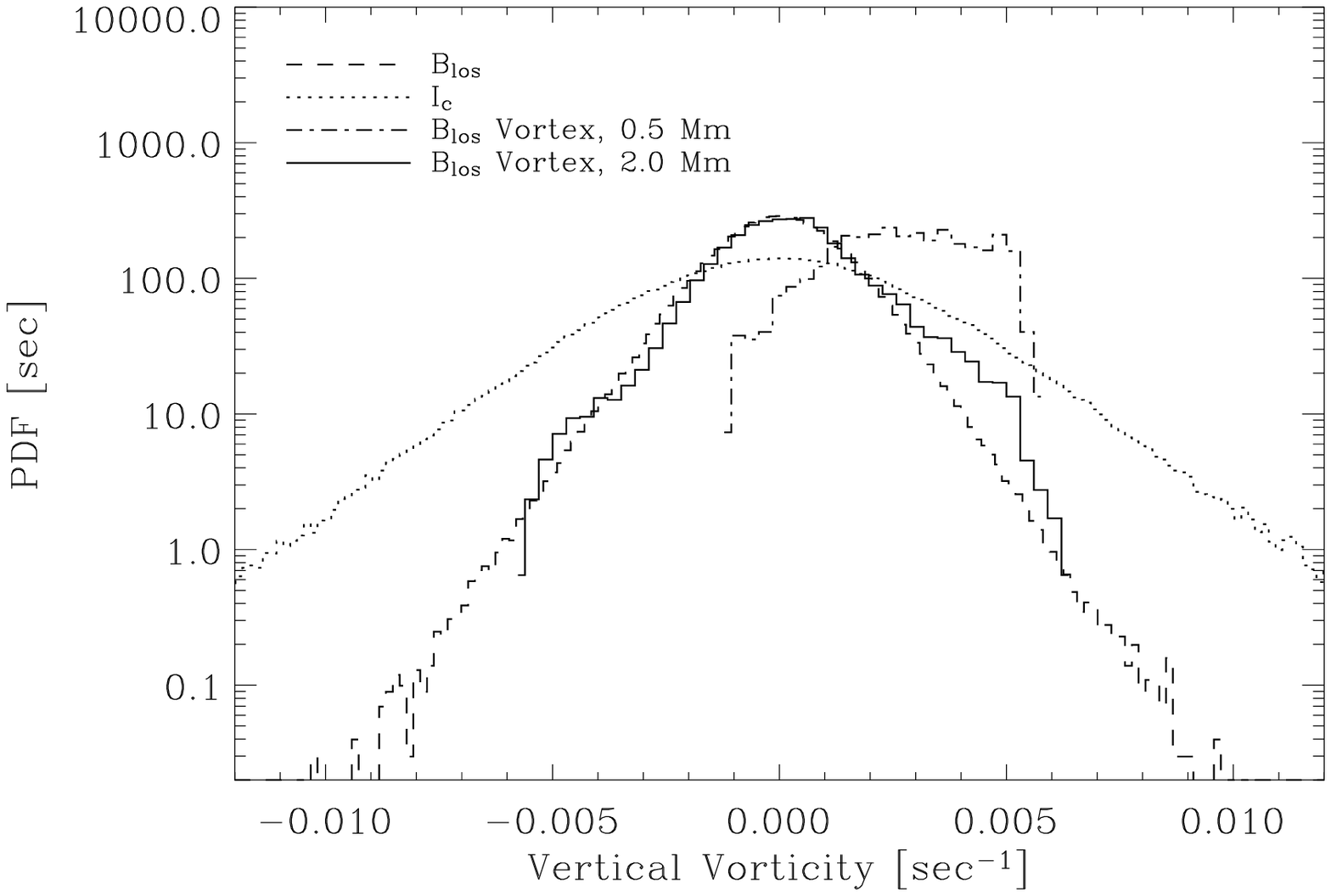}
\caption{Probability density function (PDF)
of vertical vorticities obtained from 
LCT proper motions. The 5\,min time 
averaged distributions for the full FOV are shown as 
a dotted line, and a dashed line, 
depending on whether they were derived from the 
continuum intensity or the magnetogram,
respectively. 
The dotted-dashed line corresponds to the 
vorticity in a region 0.5~Mm wide around a number 
of well defined vortices (with the vorticity signed
so that all vortices have positive vorticity). 
The solid line also represents a local histogram 
considering a region of 2~Mm. The excess of
vorticity at some 0.005~sec$^{-1}$ is produced
by the vortices. 
}
\label{histograms}
\end{figure}
Figure~\ref{histograms} shows histograms of
vertical vorticities to characterize our 
measurements. 
The solid line corresponds to the 
vorticity in a region 2~Mm 
wide around a number 
of well defined vortices.
The histogram considering only 0.5~Mm
is shown as the dotted-dashed line.
These histograms reveal the signature of
the vortices, which turn out to
have vorticities up to 0.006~sec$^{-1}$,
corresponding  
to a period of rotation of some 35\,min. 
These are vorticities inferred from the magnetograms, which 
are systematically smaller than those obtained from the 
other observables. The difference
can be pinned down to the proper motion velocity field, 
which tends to zero outside the large magnetic concentrations, 
where the polarization signals are low. The proper motions 
inferred from the other parameters extend throughout
(Fig.~\ref{fig2}), 
so that the histograms of vorticities have less contribution
at low values and show extended tails 
(cf. the dashed line and the dotted line in Fig.~\ref{histograms},
which represent the histograms of vorticities for the 
full FOV inferred from the magnetogram and the 
continuum intensity, respectively).

If a vortex has its axis tilted with respect to the \los ,
it should produce a characteristic Doppler signal similar
to the rotation curve of a galaxy, with a close pair
redshift-blueshift centered at the sinkhole. 
The expected signals are of the order of a few hundred 
m\,s$^{-1}$ for moderate-high inclinations (30\degr ), 
which are at the limit of our observation.
We unsuccessfully seek for such signals in the Doppler maps, 
meaning that the vortex motions are not highly tilted
with respect to the horizontal plane. Moreover, we note the 
discovery of horizontal vortex flows near the edges of granules 
reported by \citet{ste1}  using the same {\sc Sunrise}/IMaX data and that are 
likely to be of a different nature to those analyzed here.

%
\section{Discussion}\label{discussion}

Small-scale vortex flows in quiet Sun are detected using 
five different observational parameters: magnetogram, continuum 
intensity, line core intensity,
LOS velocity, and line width. The fact that 
in most cases the detection is consistent in three or more 
of these observables (showing different morphology) 
reinforces the reliability of the events found.

The number density of vortices is $\sim$1.7 times 
larger than that found by \citet{bon08}, even though 
we have been far more strict here. The increase
can be ascribed to the use of a larger variety 
of physical parameters to detect the swirls. 
Most of the vortices are shortlived events
observed during less than 10 min,  but some of them
last longer than the full time series, with recurrent 
vortices appearing in roughly the same place.

Vortices have a typical vorticity smaller than 
6$\times$10$^{-3}\,$sec$^{-1}$, 
which corresponds to a period of rotation of some 35\,min.
For reference, the large maelstrom found by 
\citet{bra88} had a vorticity ten times smaller,
with an associated period of some 6 hours.
The measured vorticities are generally much smaller
than those predicted by the numerical simulations
of magneto convection 
\citep[][Stein 2010 private communication]{stei98}.
We think that the bulk of this difference
can be attributed to the limited spatial-temporal
resolution of the observations. We are 
unable to identify vortices smaller than the tracking
window, and/or lasting less than 8-10 frames, which
sets an upper limit to the vorticity of some 0.04~sec$^{-1}$. 
Simulations indicate that the highest vorticities 
occur at the smallest resolvable scales and, thus, 
the predicted distribution critically depends on the
numerical resolution of the simulation
\citep[see Fig.~31 in][]{stei98}. 

The curvature maps (i.e., maps of the inverse radius of 
curvature) show the presence of vortices much better than 
the vorticity maps (see Figs.~\ref{fig3},~\ref{fig4}). The vorticity is sensitive 
to the flow speeds, which are large outside vortices, 
creating spurious vorticity signals. The curvature, however, only 
enhances areas where the swirling motions occur at 
small scales, independently of the velocity. 
The sinkholes show up conspicuously as 
local extremes in the curvature maps, and this 
new property should be employed when 
devising automatic algorithms to detect vortices. 

We find a preferred sense of rotation for the vortices 
(27 counterclockwise vs 15 clockwise).
If the two senses of rotation were equally-probable
then our observation would be highly unlikely (the probability is 4.4\% 
assuming a binomial distribution). Nevertheless, the statistics is not 
large enough to provide a firm conclusion.
 The role of Coriolis forces 
on setting up this difference can also be discarded since 
the vortex motions involve time scales much too short 
to be affected by the solar rotation. 
The preferred sense of rotation 
may have to do with the solar differential rotation. 
The plasma poleward from the sink tends to 
lag behind, whereas the plasma equatorward from the sink
moves forward.
Such a difference impinges a preferred 
counterclockwise sense of rotation in the northern hemisphere, 
and a clockwise sense in the southern hemisphere. 
Back-of-the-envelope estimates indicate that the effect 
produces the right order of magnitude 
vorticity.\footnote{\label{foot}Assume 
plasma separated by 40\,Mm (i.e., the
size of a supergranule) and converging to a sinkhole in the middle. 
Then the difference of differential rotations at latitude 30~deg
produces a difference of velocities of some 35~m\,sec$^{-1}$.
If the circulation is approximately conserved during the 
convergence, a vortex 0.5~Mm wide would have a vorticity
of 0.01~sec$^{-1}$. 
}
If this conjecture turns out to be correct, 
it naturally explains the  difference with respect to 
\citet{bon08}, whose observations correspond to the solar
equator where there is no preferred sense of rotation. 
In our case the dominant counterclockwise rotation is consistent with 
an observed FOV in the northern hemisphere.

%
%
\acknowledgements

Thanks are due to R. Stein for discussions on the 
comparison with numerical simulations, and to C. Pastor for 
her support during the data interpretation. 
The German contribution to SUNRISE is funded by the 
Bundesministerium f\"ur Wirtschaft und Technologie through
Deutsches Zentrum f\"ur Luft- und Raumfahrt e.V. (DLR), Grant No. 50 OU 0401, and by the
Innovationsfond of the
President of the Max Planck Society (MPG). 
The Spanish contribution has been funded by the Spanish MICINN under
projects ESP2006-13030-C06, AYA2009-14105-C06 (including European FEDER funds),
AYA2007-66502, AYA2007-63881 and by the EC (SOLAIRE Network -- MTRN-CT-2006-035484).
The HAO contribution
was partly funded through NASA grant number NNX08AH38G. 
This work has been partly supported by the WCU grant (No R31-10016) 
funded by the Korean Ministry of Education, Science \& Technology
%
%
%
{\it Facilities:} \facility{SUNRISE/IMAX}


%

\end{document}